\documentclass[]{aa}  % LaTeX A&A  Monotype Times Fonts
\usepackage{psfig}

\newcommand{\bcma}{$\beta$~CMa}
\newcommand{\acma}{$\alpha$~CMa}
\newcommand{\ecma}{$\epsilon$~CMa}

\newcommand{\cmd}{cm$^{-2}$}
\newcommand{\cmt}{cm$^{-3}$}
\newcommand{\kms}{km\,s$^{-1}$}
\newcommand{\dens}{$n_{\rm e}$}

\newcommand{\hei}  {\mbox{He\,{\sc i}}}
\newcommand{\hi}  {\mbox{H\,{\sc i}}}

\newcommand{\di}  {\mbox{D\,{\sc i}}}

\newcommand{\niti} {\mbox{N\,{\sc i}}}
\newcommand{\nitv} {\mbox{N\,{\sc v}}}
\newcommand{\oii}  {\mbox{O\,{\sc ii}}}
\newcommand{\oi} {\mbox{O\,{\sc i}}}
\newcommand{\ovi}  {\mbox{O\,{\sc vi}}}
\newcommand{\nai}  {\mbox{Na\,{\sc i}}}

\newcommand{\caii} {\mbox{Ca\,{\sc ii}}}

\newcommand{\cii}  {\mbox{C\,{\sc ii}}}

\newcommand{\civ}  {\mbox{C\,{\sc iv}}}

\newcommand{\nv}   {\mbox{N\,{\sc v}}}
\newcommand{\mgii} {\mbox{Mg\,{\sc ii}}}

\newcommand{\mgi}  {\mbox{Mg\,{\sc i}}}
\newcommand{\silii} {\mbox{Si\,{\sc ii}}}
\newcommand{\siliii}{\mbox{Si\,{\sc iii}}}
\newcommand{\siliv} {\mbox{Si\,{\sc iv}}}

\newcommand{\sii}{\mbox{S\,{\sc ii}}}

\newcommand{\siii} {\mbox{S\,{\sc iii}}}

\newcommand{\feii} {\mbox{Fe\,{\sc ii}}}

\newcommand{\ari } {\mbox{Ar\,{\sc i}}}

\begin{document}

   \thesaurus{ 10.19.1, % solar neighbourhood
               09.03.1,  % ISM: clouds
               09.19.1, % ISM: structure
               08.09.2 \ecma,  % stars: \ecma
               13.21.3)  % UV : ISM
               }
   \title{Local clouds: ionization, temperatures, electron densities and 
   interfaces, from GHRS and IMAPS spectra of 
   $\epsilon$ Canis Majoris
\thanks{Based on observations with the 
NASA/ESA \it Hubble Space Telescope, \rm obtained at the Space Telescope
 Institute, which is operated by the Association of Universities for 
Research in Astronomy, Inc., under NASA contract NAS5-26555.}}

   \author{C\'ecile Gry$^{1,2}$ and Edward B. Jenkins$^{3}$}

   \offprints{C. Gry}

\institute{$^{1}$ ISO Data Center, ESA Astrophysics
              Division, PO box 50727, 28080 Madrid, Spain (present address)
              (cgry@iso.vilspa.esa.es)\\
              $^{2}$ Laboratoire d'Astronomie Spatiale,
              B.P.8,
              13376 Marseille cedex 12,
              France\\
              $^{3}$ Princeton University Observatory,
              Princeton, NJ 08544-1001, USA
              (ebj@astro.princeton.edu)
}

   \date{Received 20 September 2000 ; accepted 7 December 2000}
   \authorrunning{Gry \& Jenkins}
   \titlerunning{The local clouds toward $\epsilon$~CMa}

   \maketitle

   \begin{abstract}
The composition and physical properties of several local clouds, including the
Local Interstellar Cloud (LIC) 
in which the Sun is embedded, are derived from absorption features
in the UV spectrum of the star \ecma. \\
We derive temperatures and densities 
for three components by combining our interpretations of
the ionization balance of magnesium and
the relative population of \cii\ in an excited fine-structure level.
We find that for the LIC  
\dens\~=~$0.12\pm0.05$~\cmt\ and T~=~7000$\pm1200$~K.\\
We derive the ionization fractions of hydrogen and discuss
the ionizing processes. In particular the hydrogen  and helium
ionizations in the LIC 
are compatible with  photoionization by the local EUV radiation fields
from the hot stars and the cloud interface with the hot gas.\\
We confirm the detection of high ionization species ~:  
\siliii\ is detected in all clouds and \civ\ in two of them, including the LIC,
suggesting the presence of
ionized interfaces around the local clouds.

   \end{abstract}
\section{Introduction}\label{sec:intro}
The star \ecma\
(B2~II, $V=1.50$, $v\sin i= 44\,$\kms, $\ell=239.8$, $b=-11.3$)
(Hoffleit \& Jaschek 1982) at a distance
of 130~pc (Perryman et al. 1997) is by far the brightest extreme ultraviolet 
source in the sky
(Vallerga et al. 1993) and thus the main photoionization source
in the Solar Neighborhood (Vallerga and Welsh 1995, Vallerga 1998). 
This is principally due to the 
extraordinary emptiness of the line of sight to \ecma, as we describe below.
A beneficial aspect of the emptiness and hence the simplicity
of the sight line is the opportunity for us to study with an
unusual level of detail individual diffuse clouds in the local
interstellar medium. This is the subject of this paper.

In particular, the star \ecma\ provides an opportunity to observe the absorption
spectrum of the Local Interstellar Cloud (LIC)
surrounding our solar system.  In most cases we obtained
a good signal-to-noise ratio for the absorption features, thereby 
allowing us to derive
the chemical and physical properties within the clouds,
including depletion, temperature,
electron density and ionization.

The temperature of the LIC is usually derived from the width
of absorption lines by profile fitting. Different measurements applied to
various elements all point to a temperature of around 7000~K, but with a 
relatively large error. The most precise measurements are
provided by
the observation of the HI L$\alpha$ line, because the low mass of 
hydrogen offers the best discrimination between thermal and turbulent
line broadening when compared to the results from heavier elements. Using
this method, Linsky et al. (1995) found T=7000$\pm$900\,K in the directions
toward Capella and Procyon.

The electron density in the LIC has been determined in the 
lines of sight towards Sirius (Lallement et al. 1994) and \ecma\ 
(Gry et al 1995) using the ratio $N$(\mgii)/$N$(\mgi), towards $\delta$ Cas
with the ratio $N$(\nai)/$N$(\caii) (Lallement and Ferlet 1997), and
with the ratio $N$(\cii *)/$N$(\cii) towards Capella (Wood and Linsky 1997)
and the white dwarf REJ1032+532 (Holberg et al. 1999).
All methods give results that are around 0.1 \cmt, which is roughly the same 
order of magnitude as the neutral gas density.

There are strong indications that the LIC and
similar clouds in the Local Interstellar Medium (LISM)
are partly ionized.
The observed fractional ionization of hydrogen can be explained
by the EUV radiation from white dwarf and other stars, in particular 
\bcma\ and \ecma\ (Vallerga 1998).
Previous studies of the nearby line of sight toward \bcma\ with 
the {\it Goddard High Resolution Spectrograph\/} (GHRS) on the Hubble Space
Telescope (Dupin and Gry 1998) and an independent facility, the {\it Interstellar
Medium Absorption Profile Spectrograph\/} (IMAPS) (Jenkins, Gry and Dupin 2000)
have shown that the two main clouds 
in that sight-line present a  very high ionization fraction of hydrogen, 
which could be  explained by photoionization due to
the combination of \ecma\ and \bcma\ if the clouds are located close
enough to the stars.

An outstanding problem is that these and other stars do not
produce enough photons with
energies above 24.6$\,$eV to explain the high fractional ionization of helium 
in the LISM, as shown by the low value of n(\hei)/n(\hi) (equal
to 0.07 instead of the cosmic ratio of 0.1, which indicates that
helium is more ionized than hydrogen). 
The two main proposals to explain this phenomenon are {\it i)}
the LISM is still recombining from
a much more 
highly ionized state produced by a supernova-related energetic event 
in the recent past (Reynolds 1986 ; Frisch \& Slavin 1996 ; Lyu \& Bruhweiler
1996) and {\it ii)} the ionization of He is maintained by the diffuse
EUV radiation emitted by conductive interfaces between the cloud edges 
and the hot gas filling the ``Local Bubble'' in which they are embedded
(Slavin 1989 ; Slavin \& Frisch 1998).

In this paper we present GHRS spectra of  
\ecma\ over limited wavelength intervals between 1190 \AA\ and 1550 \AA, ones that
include the lines of \niti, \oi,
\cii\ and \cii*, \sii, \silii, \siliii, and \civ\ at a wavelength resolution 
R\,$\sim$\,100\,000.  We consider also a profile
of \oi\ at 1039 \AA, recorded by 
IMAPS at R\,$\sim$\,60\,000. We
show how they shed light on the  knowledge of characteristics of 
the nearby  diffuse clouds such as temperatures, electron densities,
abundances, and degree of ionization.

\section{Observations and data reduction}\label{sec:data}
Most of the observations presented here have been performed in late 1996 with 
the Ech-A grating of GHRS.
All data were taken with the 0$\farcs$25 Small Science Aperture 
(SSA), the procedure FP-SPLIT = 4 and a substepping of 4 samples per diode 
(for details of the instrumentation, see Soderblom et al. 1995). 
For data processing, we used the standard STSDAS procedures working in the 
IRAF environment. We assigned wavelengths from the standard 
calibration tables. An error of $\pm$1 resolution element on the wavelength 
assignment is expected to arise from magnetic drifts.

The signal-to-noise ratio (S/N) for all Ech~A data  is about 200  
when the flux is at the level of the stellar continuum. However there is a
degradation of signal quality for features that appear in the bottoms of strong
stellar lines. For most interstellar
lines the S/N ranges between 100 and 200, but it is 80 for \cii\ and
\cii*\, and 30 in 
the extreme case of \siliii\, where the stellar line is the deepest.

Observations in the far-UV lines were carried out 
by IMAPS when it was operated on the
ORFEUS-SPAS~II mission that flew in
late 1996 (Hurwitz, et al. 1998).  IMAPS is an objective-grating echelle
spectrograph that was designed to record the spectra of bright,
early-type stars over the wavelengths from $\sim950$~\AA\ to
$\sim1150$~\AA\ with a high spectral resolution. For more details on the
instrument see Jenkins et al. (1996).

The IMAPS spectra were extracted from the echelle spectral images using
special procedures developed by one of us (EBJ) and his collaborators on
the IMAPS investigation team. The S/N obtained
for the interstellar lines observed by IMAPS is on the order of 25 to 30.
 
In both sets of spectra, there is an uncertainty related to the
background correction for scattered light on the echelle format.
For the GHRS Ech-A data, the error in the background correction
is smaller than a few percent of the continuum and is only a concern for  
strong lines that almost reach the zero intensity level in their deepest points.
In the case of a species for which several lines with different
oscillator strengths are available, such as \silii , the uncertainty 
is eliminated by comparing the different profiles, which must have relative 
strengths that follow their oscillator strengths. 
For very strong lines like \hi\ Ly$\alpha$ and \cii\ 1334\AA,
the zero level is determined by the base of the saturated line. 
The situation for the \oi\ 1302\AA\ line is reported in
Section~\ref{sec:oicolumn}.
\begin{figure}
\psfig{file=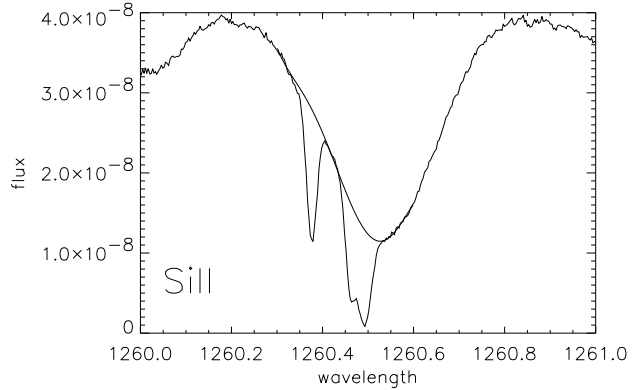,width=8cm}
\caption{Example of the continuum normalization for the \silii\ 
line at 1260.4\AA. The stellar line is fitted by a polynomial (shown here 
superimposed on the spectrum) over the velocity
interval [-30\kms, 40\kms].} 
\label{fig:stellar}
\end{figure}

Stellar lines were 
fitted with one low-order polynomial over a velocity range covering the  
interstellar components (\S\ref{sec:column}), by considering the points 
bluewards of Component~4 up to V=-30 \kms\ or -40 \kms,
redwards of Component~1 up to V=+40 \kms, as well as the velocity range 
which is not affected by any interstellar absorption between 
Components~2 and 3. The spectra are then
normalized to these stellar lines to create the interstellar profiles
with a level continuum. An example of the stellar line 
fitting is illustrated in Figure~\ref{fig:stellar} for the 
\silii\ 1260\AA\ line.
For strong or moderately strong lines the uncertainty added to the
column density estimates by this normalization process is small.
However, for very faint lines (\mgi, \cii*, \sii), it is a major
source of error. This uncertainty was taken into account when we listed
the column density results in Table~\ref{tab:columns}. The most difficult
case is that for \sii ,  where 
all three interstellar lines are very faint and 
appear on top of a steep slope of a stellar line. In this 
configuration there is a lot of
freedom for the location of the synthetic stellar lines which  
can artificially enhance the 
interstellar line or, alternatively, make it disappear almost completely.
We thus use several options for the \sii\ absorption profiles defined
by an envelope of $\pm2\sigma$ around the fitted continuum.

We derive the column densities by using the line fitting software 'Owens'
developed by Martin Lemoine. Each interstellar absorption component is 
represented by the convolution 
of a theoretical Voigt profile with the instrumental profile. The instrumental
profile for the GHRS Echelle data was assumed to be a Gaussian 
with a FWHM of 0.92 diodes 
(Soderblom et al. 1995). 
An iterative 
procedure which minimizes the sum of the squared differences between
model profiles and 
the data points allows us to determine the most likely column densities of the 
absorbing elements {\it N}(cm$^{-2}$), the radial velocity of the cloud (\kms) 
and the velocity dispersion ($b$-value) (\kms) of each interstellar absorption
component. The
software also allows us to fit the lines from several elements simultaneously,
which leads directly to a coherent solution for all species in
terms of velocity, temperature and turbulent velocity.
The wavelengths and $f$-values are listed 
in Table~\ref{raiestud}. 
\begin{table}
\caption[]{\label{raiestud}Observed atomic lines.$^1$}
\begin{tabular}{cccc}
\hline
\noalign{\smallskip}
&Element & wavelength (\AA) & $f$-value\\
\noalign{\smallskip}
\hline
\noalign{\smallskip}
IMAPS&\oi & 1039.230 & $9.20\,10^{-3}$\\
Ech A&\siii & 1190.203 & $2.31\,10^{-2}$\\
&\silii & 1190.416 & $2.93\,10^{-1}$\\
&\silii & 1193.290 & $5.85\,10^{-1}$\\
&\niti & 1199.550 & $1.30\,10^{-1}$\\
&\niti & 1200.223 & $8.62\,10^{-2}$\\
&\niti & 1200.710 & $4.30\,10^{-2}$\\
&\siliii & 1206.500 & $1.67\,10^{0}$\\
&\di & 1215.339 & $4.17\,10^{-1}$\\
&\hi & 1215.670 & $4.17\,10^{-1}$\\
&\sii & 1250.584 & $5.43\,10^{-3}$\\
&\sii & 1253.811 & $1.09\,10^{-2}$\\
&\sii & 1259.519 & $1.66\,10^{-2}$\\
&\silii & 1260.422 & $1.18\,10^{0}$\\
&\oi & 1302.168 & $5.19\,10^{-2}$\\
&\silii & 1304.370 & $9.17\,10^{-2}$\\
&\cii & 1334.532 & $1.28\,10^{-1}$\\
&\cii* & 1335.708 & $1.15\,10^{-1}$\\
&\siliv & 1393.755 & $5.14\,10^{-1}$\\
&\siliv & 1402.770 & $2.55\,10^{-1}$\\
&\civ & 1548.195 & $1.90\,10^{-1}$\\
&\civ & 1550.770 & $9.48\,10^{-2}$\\
Ech B&\mgii & 2803.503 & $3.06\,10^{-1}$\\
&\mgii & 2796.352 & $6.15\,10^{-1}$\\
&\mgi & 2852.964 & $1.83\,10^{0}$\\
&\feii &2344.214 & $1.14\,10^{-1}$\\
&\feii &2382.765 & $3.20\,10^{-1}$\\
&\feii &2586.650 & $6.91\,10^{-2}$\\
\noalign{\smallskip}
\hline
\end{tabular}\\
\footnotesize{
$^{1}$ Wavelengths and $f$-values are from a private communication by Morton,
updating data from Morton (1991).
}
\end{table}
\section{Line of sight structure and column densities}\label{sec:column}
\begin{table*}
\caption[]
{Column densities (cm$^{-2}$) of the five components 
detected toward \ecma.}
%\scriptsize{
\begin{tabular}{lccccc}
\hline
\noalign{\smallskip}
Comp.&  1 & 2 & 3 & 4 & 5\\
$V$ & 17 \kms & 10 \kms & $-10$ \kms & $-19$ \kms & $-65$ \kms \\
\noalign{\smallskip}
\hline
\noalign{\smallskip}
\feii   & $1.35\pm0.05\,10^{12}$ &
        $5.2\pm0.5\,10^{11}$ & $1.99\pm0.1\,10^{11}$ & - & - \\
\noalign{\smallskip}
\mgii  & $3.1\pm0.1\,10^{12}$ & $1.05\pm0.05\,10^{12}$
         & $1.39\pm0.05\,10^{12}$ & $9.0\pm4\,10^{10}$ & -\\
\noalign{\smallskip}
\mgi    & $7\pm2\,10^{9}$ & $5\pm2\,10^{9}$ &
        $7\pm3\,10^{9}$ & - & -\\
\noalign{\smallskip}
\silii    & $4.52\pm0.2\,10^{12}$ &
        $1.85\pm0.1\,10^{12}$ & $1.38\pm0.05\,10^{12}$ &  
        $2.9\pm2\,10^{10}$ & -\\
\noalign{\smallskip}
\siliii   & $2.3\pm0.2\,10^{12}$ & $2.0\pm1.1\,10^{11}$ &
        $1.0\pm0.1\,10^{12}$ & $2.2\pm0.5\,10^{11}$ & $2.1\pm0.3\,10^{11}$ \\
\noalign{\smallskip}
\cii\ $^a$   & $(1.4 - 2.1)\,10^{14}$ &
        $(0.7 - 1.3)\,10^{14}$ & $5.21\pm0.2\,10^{13}$ & $7.3\pm2\,10^{12}$ &
        $2.3\pm0.3\,10^{12}$ \\
\noalign{\smallskip}
\cii*    & $1.3\pm0.2\,10^{12}$ & $2.0\pm1.1\,10^{11}$ &
        $7.4\pm1.0\,10^{11}$ & - & -\\
\noalign{\smallskip}
\oi    & $1.4^{+0.5}_{-0.2}\,10^{14}$ &
        $1.2\pm0.3\,10^{14}$ & $1.93\pm0.2\,10^{12}$ & - & - \\
\noalign{\smallskip}
\niti    & $1.70\pm0.05\,10^{13}$ &
        $9.8\pm0.5\,10^{12}$ & $<2.2\,10^{11}$ & - & -\\
\noalign{\smallskip}
\sii    & $8.6\pm2.1\,10^{12}$ & $4.9\pm1.5\,10^{12}$ &
        $3.5\pm0.9\,10^{12}$ & - & - \\
\noalign{\smallskip}
\civ  & $1.2\pm0.3\,10^{12}$ & - &
        $3.7\pm0.4\,10^{12}$ & - & - \\
\noalign{\smallskip}
\siii   & $<3\,10^{12}$ & $<3\,10^{12}$ & $<3\,10^{12}$
        & - & - \\
\noalign{\smallskip}
\siliv    & $<2\,10^{11}$ & $<2\,10^{11}$ &
     $<2\,10^{11}$ & - & -\\
\noalign{\smallskip}
\nv    & $<2.5\,10^{11}$ & $<2.5\,10^{11}$ &
$<2.5\,10^{11}$
        & - & -\\
\noalign{\smallskip}
\hline
\end{tabular}
\\
\footnotesize{
$^a$ the upper limits for Components~1 and 2 are determined from N(\sii) 
as explained in Section~\ref{sec:ciicolumn}}
\label{tab:columns}
\end{table*}
Previous high resolution 
spectra of UV absorption lines 
for the 
elements \feii, \mgii\ and \mgi\ obtained with the GHRS Ech B revealed a simple
structure
for the interstellar medium in the line of sight toward \ecma\ (Gry et al.
1995) : only three  components were detected in all absorption lines, and
their contributions amounted to an equivalent hydrogen column density of 
less than $10^{18}{\rm cm}^{-2}$, a value that
is by far the lowest amount ever observed in the Galactic
disk for a line of sight as long as 130~pc. The low column densities
offer the advantage
that most of the UV absorption lines are unsaturated and can yield reliable
column densities. Small, extra velocity components are detected 
in some of the strongest lines, as shown by Gry et al. (1995), however none 
of these exceeds an equivalent H column density of about $4\,10^{16}{\rm
cm}^{-2}$.

The three principal components had heliocentric velocities of 17, 10 and $-$10
\kms. In the previous papers, they were identified as Components~1, 2 and 3,
respectively.  Two of the components have been identified by Gry et al (1995) as
the two clouds detected in the line of sight to Sirius by Lallement et al. (1994)
and described further by H\'ebrard et al. (1999). As Sirius is located at a 
distance of only 2.7 pc, these two components must be situated very close to 
the Sun.

Component~1 is recognized as the Local Interstellar
Cloud (LIC) in which the Sun is embedded, for which the motion has been 
characterized by Lallement and Bertin (1992).  Component~2, called the ``Blue
Cloud,'' is also in front of Sirius but is distinct from the LIC. Another
component (Component~4) was detected on the 
blue side of Component~3 in the
strongest lines at $-$19 \kms, while a fifth component was detected only in the 
line of \siliii\ at $-$65 \kms, and confirmed in \hi\ because it
mimics a very strong \di\ absorption (too strong to be due to \di\ with a 
reasonable D/H ratio), for
its \hi\ absorption turns out
to coincide with the expected \di\ feature from Component~1 
(see \S\ref{sec:hidi}). Finally,
a ``Component~0'' had been  introduced in the red wing of Component~1 to improve 
the fit of some of the profiles, but our reanalysis with more complete data
indicates that this  component may not be real. Its introduction
in the previous Ech~B data analysis was probably a consequence of using
a slightly distorted line spread function. We have chosen to omit the
component in our more refined analysis.
\begin{figure*}
\psrotatefirst
\psfull
\psfig{file=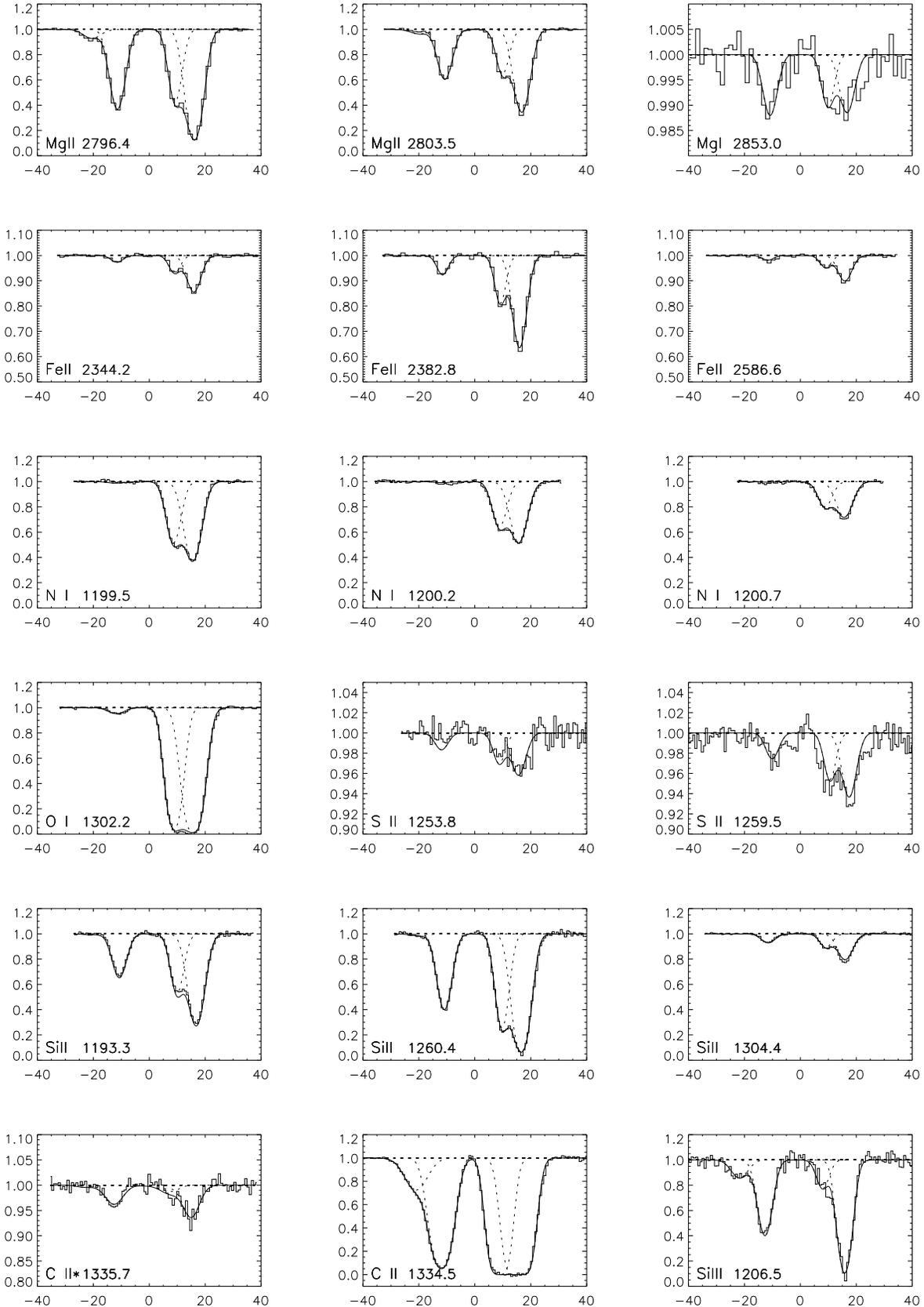,height=23cm}
\caption{Ech-B (first six plots) and new Ech-A spectra of \ecma\ 
($R\sim100\,000$), plotted in the heliocentric velocity scale (\kms). 
The histogram-style
tracings represent the observations, and the smooth solid
lines indicate the fits to the interstellar absorption profiles. 
Dotted lines indicate the assumed stellar 
continua and the individual component contributions. The spectra have 
previously been divided out by a stellar
absorption profile, as explained in Section~\protect\ref{sec:data}. 
}
\label{fig:fits}
\end{figure*}

We have performed the line fitting for Components~1 to 4 over
all elements available in the Echelle~A
and Echelle~B data, as well as for the \oi\ 
line recorded by IMAPS. 
All lines of \oi, \niti, \mgi, \mgii, \feii,
\silii, \siliii, \cii, \cii**, and \sii\ were fitted simultaneously
with a unique line-of-sight
velocity structure and a unique column density that was consistent 
with different lines of the same species. As the absolute wavelength 
calibration
had a precision of one resolution element (about 3 \kms),
an individual velocity shift is allowed for each data set, whereas the 
relative velocity of the components had to be the same for all lines.
The resulting velocity shifts for all elements have a dispersion of
$\pm0.94$~\kms, in agreement with the  precision of the wavelength calibration.
The derived velocity for Component~1 is 16.15$\pm1.5$~\kms, in agreement
with its identification as the Local Interstellar Cloud.
The velocity shifts of the other components relative to Component~1
are known with more precision : $-6.92\pm0.03$~\kms\ for Component~2, 
$-27.50\pm0.04$~\kms\ for Component~3, and
$-36.4\pm0.3$~\kms\ for Component~4.

The comparison of the synthetic profiles with the data
suggested that the velocity shift between Component~3 and Component~1 
could be slightly different for \siliii\ from what we found for 
the other elements. We thus decided to decouple the \siliii\ velocity 
from the determinations for other species,
and we found that indeed the velocity shift
between Component~3 and Component~1 for \siliii\ is $-28.48$~\kms, i.e.,  
1~\kms more than for the other elements. This velocity
shift however is only one-third of a resolution element and must be compared to
the velocity dispersion of material in the clouds which is more than
a factor of three higher.
\begin{table}
\caption[]
{b-values (velocity dispersion in \kms) derived from the fits. They are 
given for the elements carbon and iron, which represent the two extremes of atomic
mass. The $b$-values for the other elements are intermediate between the two.
}
\begin{tabular}{lcccc}
\hline
\noalign{\smallskip}
Comp.&  1 & 2 & 3 & 4 \\
\noalign{\smallskip}
\hline
\noalign{\smallskip}
\cii & $3.9\pm0.3$&$3.2\pm0.13$&$4.0\pm0.1$&$5.5\pm0.1$\\
\noalign{\smallskip}
\feii   & $3.0\pm0.15$&$2.12\pm0.17$&$2.44\pm0.02$&$4.4\pm0.4$\\
\noalign{\smallskip}
\hline
\end{tabular}
\label{tab:bvalues}
\end{table}

When we derived $b$-values that had the best fit to the data, we imposed the
requirement that various elements should have common values for the turbulent and
thermal doppler contributions within each velocity component.  Differences in the
outcomes for $b$ across different elements were therefore governed only by
variations in atomic mass.
Table~\ref{tab:bvalues} shows the $b$-values for two elements that represent the
two extremes in mass: carbon and iron.

The derived column densities for Components~1 to 4 
are listed in Table~\ref{tab:columns}, and the fits of the
lines are presented in Figure~\ref{fig:fits}.
Specific considerations for a few elements are
discussed individually below.
\begin{figure}[h!tb]
\psrotatefirst
\psfull
\psfig{file=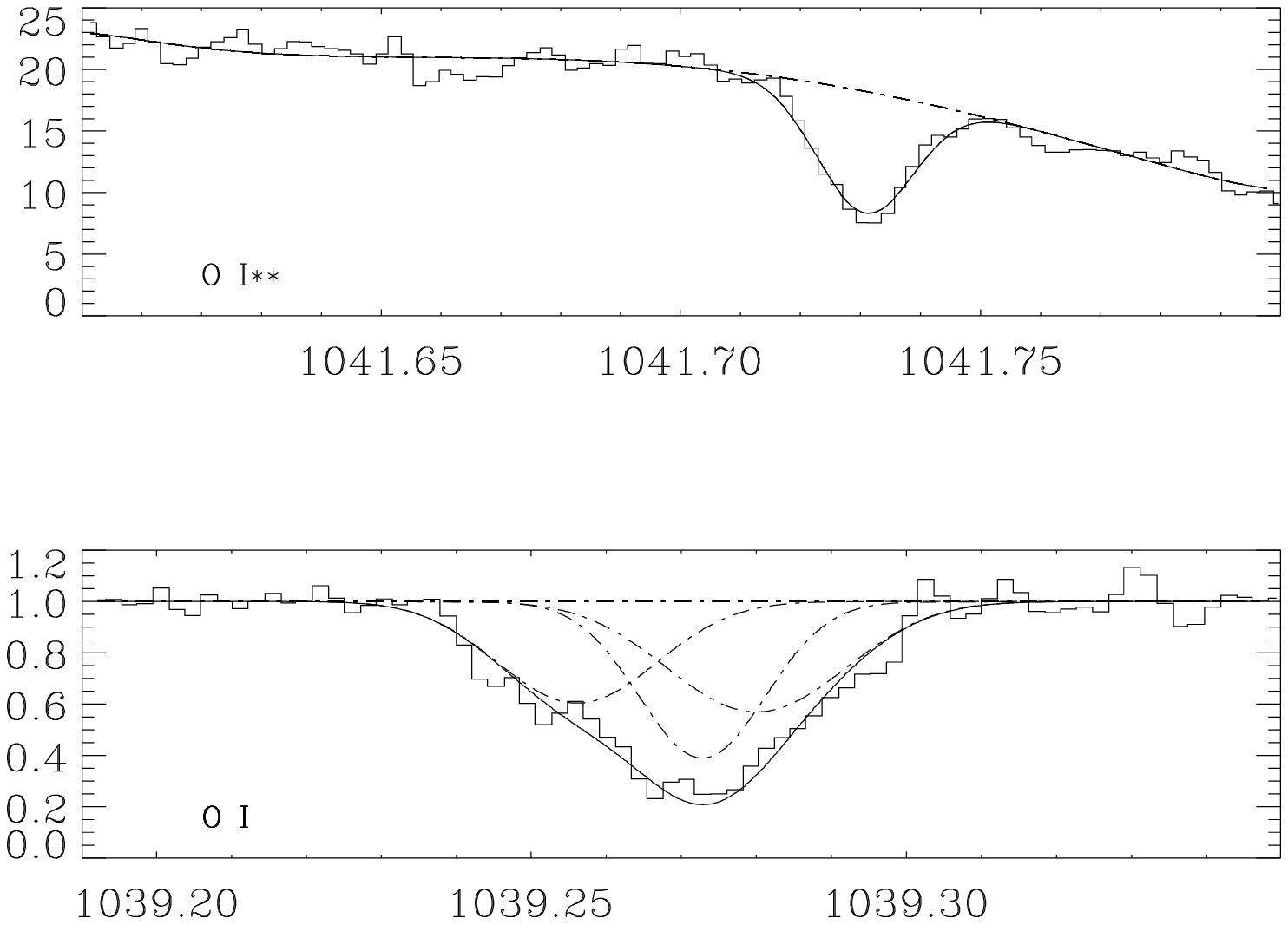,width=8cm}
\caption{IMAPS spectra of \ecma\ ($R\sim60\,000$) for the line of \oi\ at
1039\AA\ and
\oi** at 1041\AA. The histogram-style
tracings represent the observations and the smooth solid
lines the fits to the interstellar and telluric absorption profiles. 
Also shown in dash-dotted lines are the assumed stellar 
continuum and the individual component contributions.
In the \oi** line, only the telluric component contributes to the profile 
while the \oi\ line is a blend of the interstellar components 1 and 2 and
the telluric line (the telluric component is the central one).
}
\label{fig:oi}
\end{figure}
\subsection{\oi\ column density}\label{sec:oicolumn}
Guidance on the correct zero level of the \oi\ line  at 1302\AA\ was available
from the observed zero level of the nearby \cii\ line at 1334\AA\ which 
is known to be heavily saturated.  The presence of clearly visible
component structures at the bottom of the line indicates that the 
profile remains above zero or, at most, it
hits zero over a range of just a few points, in accord with the 
expected noise level. However an uncertainty 
on the exact zero level location still remains, and this is manifested
as an uncertainty in the 
column densities for Components~1 and 2.
It is thus useful to check the results using the weaker line of \oi\ at 1039\AA,
which is available  in the IMAPS data. 

The interstellar \oi\ feature observed with IMAPS is unfortunately contaminated by
telluric \oi\ absorption,
as revealed by the presence of a strong, narrow \oi** absorption line 
at 1041 \AA.
However we can estimate the expected telluric contribution for \oi .  In the
Earth's upper atmosphere, the fine-structure levels
should be populated in accord with their statistical weights.
With this in mind, we performed a 
fit of the far-UV \oi\ and \oi** lines together with the other species 
(leaving out  the \oi\ 1302\AA\ line), but with the addition of a
telluric component satisfying the constraint that $N(\oi)=5\times N$(\oi**).
The resulting fits are illustrated in Figure~\ref{fig:oi}.
They give results for the \oi\ column densities of Components~1 and 2
that agree with the range of values derived from the \oi\ 1302\AA\ line
alone, giving an assurance that the  
background assignment for that line was not erroneous. 
Table~\ref{tab:columns} lists the \oi\ 
column densities which  are consistent with both \oi\ lines.
\subsection{\cii\ column density}\label{sec:ciicolumn}
There are substantial uncertainties with
the \cii\ column density determinations for 
Components~1 and 2 because the line is so strong. 
Over the velocity range covered by Component~1, the profile shows nearly zero
intensity -- a situation that is consistent with arbitrarily high column
densities. Therefore, while some lower limit for $N$(\cii) could be gathered 
from the profile fitting, we derived 
upper limits  from the upper limits for $N$(\sii) multiplied by the
cosmic abundance ratio,
$(N(\cii)/N(\sii))_{cosmic}=20$  (Anders and Grevesse,
1989) . In effect, \sii\ and \cii\ are both the dominant
ionization states in diffuse neutral gas, and even if these two elements are in a
partially ionized medium, they should have 
about the same fraction of atoms elevated to higher stages of ionization 
(Jenkins, Gry and Dupin 2000).  While this may be true, we also know that
carbon is usually more depleted
than sulfur in the ISM, implying that $N$(\cii)/$N$(\sii) is probably 
lower than the cosmic abundance ratio. The upper limits for N(\cii) listed in 
Table~\ref{tab:columns} for Components~1 and 2 reflect these considerations.
For Component~3, the line is not
as saturated, and we derived both the lower and the upper limits 
from the line fitting, and we note that the upper limit we measure
is very close to the value we would estimate from \sii.
\begin{figure}[h!tb]
\psrotatefirst
\psfull
\psfig{file=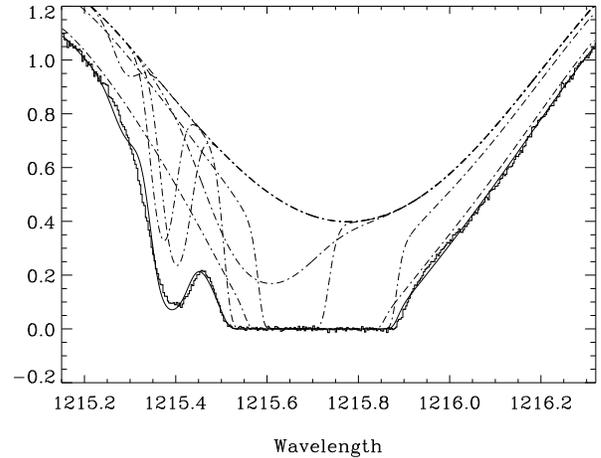,width=9cm}
\caption{The four interstellar components of \hi\ and \di\ fitted on the 
stellar Lyman $\alpha$ profile of \ecma, represented by a fourth order 
polynomial, the coefficients of which are also free parameters in the fit.
This presentation illustrates the complex superposition of profiles. 
It is 
impossible to find a unique solution for the \hi\ and \di\ column densities, 
partly because of the impossibility to define the stellar continuum 
independently.
}
\label{fig:lyman}
\end{figure}
\subsection{\hi\ and \di\ Lyman $\alpha$ profiles}\label{sec:hidi}
In principle, the low column density of neutral matter present in the line of
sight toward \ecma\ is favorable for studying D/H because
the bottom of the \hi\ line is not wide enough to bury the deuterium
line.
Indeed, as seen in Figure~\ref{fig:lyman}, the \di\ feature is clearly seen
in the wing of the \hi\ profile. 
Unfortunately, as already noted by Gry et al (1995) with the G160M data,
 it is very difficult to derive \hi\ and \di\ column densities
from the Lyman $\alpha$ line because \\
i) the \hi\ stellar line is narrow compared to the interstellar profile
and is thus very difficult to define with any accuracy, \\
ii) the \di\ feature
is probably blended with an \hi\ absorption from the small component at
$-65$ \kms.
Although this component is very weak and probably mostly ionized, for
it is detected only in \siliii\ and \cii, a neutral hydrogen
column density of only 10$^{13}$ cm$^{-2}$ could dominate over a deuterium
feature if ${\rm D/H}\sim 2~10^{-5}$. \\
iii) since the mass of deuterium is low, its $b$-value is
high and the absorption features from the different components
are not resolved very well, unlike the cases for the other elements. This makes it 
impossible to distinguish the contribution of each component to the 
\di\ profile, which might otherwise allow us to separate a blend of three or four
\di\ components from the one \hi\ high
velocity component.

We  plot in Figure~\ref{fig:lyman} some representative fits to the 
Lyman $\alpha$ profile simply to illustrate
the complexity arising from the overlapping 
individual absorbers. In this example
the Lyman $\alpha$ profile has been fitted using information from the
other elements, without any additional constraints. This solution  
is not acceptable because  it implies an \hi\ column 
density for 
Component~2 which is incompatible with the limit set by \oi\ or \sii.
We have checked nevertheless that the \hi\ column densities derived from
\oi\ in Section~\ref{sec:neutral} produce a synthetic Lyman $\alpha$ profile
which is  compatible with the observed profile. 
\subsection{\civ\ profiles}
\begin{figure}[h!tb]
\psrotatefirst
\psfull
\psfig{file=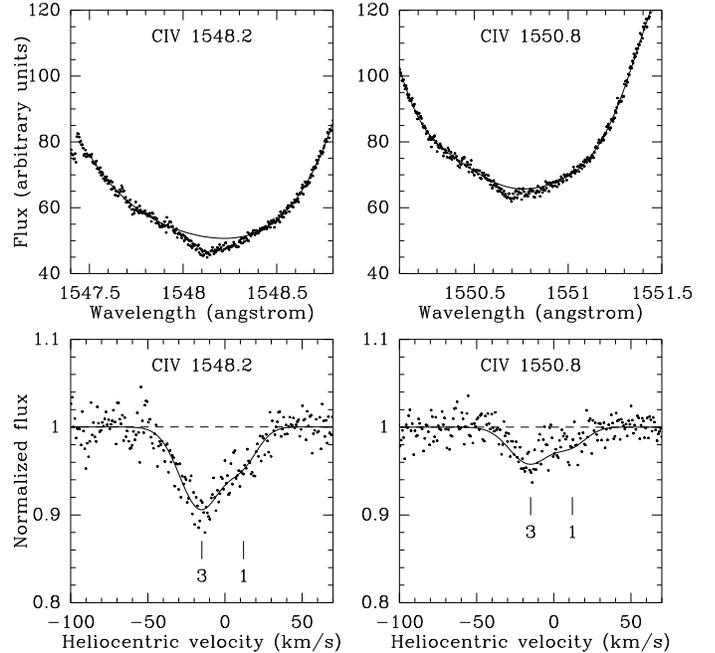,width=9cm,angle=-90}
\caption{The \civ\ profiles of \ecma\ at 1548.2 \AA\ (left) and 1550.8 \AA\
(right). This figure shows the fit of the 
stellar line by a fourth order polynomial (top), followed by the residuals
arising from interstellar absorption with two components (bottom).  
A fit (solid line)
indicates the existence of two components with a velocity separation 
equal to that of Components~1
and 3 derived from the other elements.  However the absolute velocity of \civ\ is 
shifted by $-$5~\kms compared to 
the velocities derived from the other species. The temperatures implied
by the fits are equivalent to a kinetic temperature within the range
100\,000 to 200\,000 K.
The column densities are given in Table~\protect\ref{tab:columns}
}
\label{fig:civ}
\end{figure}

Some absorption features can
be seen on the EchA spectra at the bottom of the two \civ\ stellar lines 
(Figure~\ref{fig:civ}). They were already apparent at lower resolution
in the G160M data (Gry et al. 1995).
If we fit a stellar 
continuum outside the range where the contributions from Components~1
and 3 are expected , the normalized 
spectra show two components with a velocity separation which is compatible
with the separation between Components~1 and 3 with 
column densities of 1.2$\pm0.3\,10^{12}$ and 3.7$\pm0.4\,10^{12}$~\cmd,
respectively.

There is a shift of about $-5$~\kms\ in the absolute
velocity of the components, compared to the velocities of these components
measured for the other species. This shift is a bit too high to be attributed
to an uncertainty in the wavelength calibration, which should be less than
3~\kms\ (this has been verified
with the other elements).
However, velocity discrepancies between highly ionized species and
lower ionization stages are not unusual; see, e.g., Sembach, Savage and 
Jenkins (1994).

We have also performed an evaluation of the \civ\ lines with a fit to the stellar
and interstellar lines simultaneously over a velocity interval that 
 includes the region covered by
Components~1 to 3 and without specifying the velocity of the interstellar 
components.  This approach eliminates
the possible bias that could arise from the {\it a priori} choice of the
location of the interstellar features when defining the stellar continuum. The
result of the fit
was virtually the same as that found 
with our standard way of deriving a normalized spectrum.  Once again, a velocity
separation equivalent
to that between Components~1 and 3 was found.  This separate exercise reinforces
our proposal that the \civ\ lines are real and not simply a consequence of our
incorporating information from other elements in lower stages of ionization.

\subsection{Other high ionization species}
Absorption
features from the \ovi\ doublet have been looked for in the IMAPS spectrum
but they are not visible. None of the other high ionization species  \siliv,
\siii\ and \nitv\ have been detected in the GHRS spectrum.
\section{Neutral and total gas column densities}\label{sec:hydrogen}
\subsection{Neutral gas column densities}\label{sec:neutral}
The ionization fractions of oxygen and nitrogen are coupled to that
of hydrogen via resonant charge exchange reactions.  In particular,
the rate coefficient for charge exchange of \oii\ with \hi\ is exceptionally
strong, $\sim$~$10^{-9}\,{\rm cm}^3{\rm s}^{-1}$ (Field \& Steigman 1971),
and \oi\ is therefore a very good tracer of \hi.  We can thus derive
the column densities of neutral hydrogen in the components from their respective
\oi\ column 
densities through the relation 
$N_{\rm OI}$(\hi)~=~$N$(\oi)\,/\,(O/H)$_{\rm ISM}$, where  (O/H)$_{ISM}$ is
the abundance of oxygen relative to hydrogen in the interstellar medium.
If we adopt the value derived by Meyer et al. (1998), 
(O/H)$_{\rm ISM}$=3.16\,10$^{-4}$, we find that\\
$N_{\rm OI}$(\hi)=4.4$^{+1.6}_{-0.6}$~10$^{17}$\cmd\ for Component~1,\\
$N_{\rm OI}$(\hi)=3.8$\pm$0.9~10$^{17}$\cmd\ for Component~2 and\\
$N_{\rm OI}$(\hi)=6.1$\pm$0.6~10$^{15}$\cmd\ for Component~3.\\
With these numbers, we arrive at a total \hi\ column density in the
range 7~10$^{17}$ to 1.1~10$^{18}$\cmd, which is 
compatible with that derived from the absorption of the extreme 
ultraviolet 
flux  from the star based on  EUVE spectra  (Vallerga et al. 1993 ;
Cassinelli et al. 1995).

If we adopt a different tactic by assuming that the diffuse clouds in the local
ISM are
undepleted and the oxygen abundance is equal to the
abundance in B stars, (O/H)$_*$=4.68\,10$^{-4}$, then the  \hi\
column density is lower, ranging from 4.8~10$^{17}$ to 7.2~10$^{17}$ \cmd\
for the whole line of sight, with \\
$N_{\rm OI}$(\hi)=3.0$^{+1.1}_{-0.4}$~10$^{17}$\cmd\ for Component~1,\\
$N_{\rm OI}$(\hi)=2.6$\pm$0.6~10$^{17}$\cmd\ for Component~2 and\\
$N_{\rm OI}$(\hi)=4.1$\pm$0.4~10$^{15}$\cmd\ for Component~3.

We can also derive the \hi\ column
densities implied by the \niti\ column densities. If we again use
the abundance of nitrogen in the interstellar medium derived by Meyer et al. 
(1997), (N/H)$_{ISM}$=7.5\,10$^{-5}$, we find\\
$N_{\niti}$(\hi)=2.1$\pm$0.2\,10$^{17}$\cmd\ for Component~1,\\
$N_{\niti}$(\hi)=1.3$\pm$0.1\,10$^{17}$\cmd\ for Component~2, and \\
$N_{\niti}$(\hi)$\leq$2.9\,10$^{15}$\cmd\ for Component~3. 

We note that in all three components, the neutral column density derived
from \niti\ is significantly lower than that derived from \oi.
This \niti\ deficiency  in the local ISM has already been noticed by 
Jenkins et al (2000) from  far-UV spectra 
of white dwarf stars observed with FUSE. They  have shown that  
the \niti\ deficiency, as  that of 
 \ari, favors the existence of a source of ionizing
photons with E$\geq$24.6~eV in the local ISM to explain the \hei\
ionization. \\
For the subsequent discussions, we will adopt for  N(\hi) the ranges 
derived from N(\oi), keeping both options for the oxygen depletion.
\subsection{Total gas column densities}\label{sec:totcol}
\sii\ is traditionally used as an indicator of the total 
(i.e. neutral plus ionized) gas column density in the interstellar medium because 
{\it i)} sulphur has little or no depletion onto dust grains (see, e.g.,
Savage \& Sembach (1996) and Fitzpatrick \& Spitzer (1997)) and {\it ii)} the 
ionization potential of \sii\ (23 eV) is high and therefore \sii\ is often
assumed to be the dominant ionization stage both in HI and HII regions.
If we trust that assumptions {\it i)} and {\it ii)} are valid, we can derive the
total column densities
for Components 1, 2 and 3 from the expression : 
N$_{\sii}$(H$_{tot}$)~=~N(\sii)\,/\,(S/H)$_{cosmic}$, where (S/H)$_{cosmic}$
is the cosmic abundance ratio taken from Anders \& Grevesse (1989) :
(S/H)$_{cosmic}$=1.84\,10$^{-5}$.
In this context, we obtain\\
N$_{\sii}$(H$_{tot}$)=$4.7\pm1.1~10^{17}$\cmd\ for Component~1,\\
$N_{\sii}$(H$_{tot}$)=$2.7\pm0.8~10^{17}$\cmd\ for Component~2, and\\
$N_{\sii}$(H$_{tot}$)=$2.0\pm0.5~10^{17}$\cmd\ for Component~3.
\section{Electron density and temperature}\label{sec:density}
\begin{figure*}
      \vspace{0cm}
      \hbox{\hspace{0cm}\psfig{figure=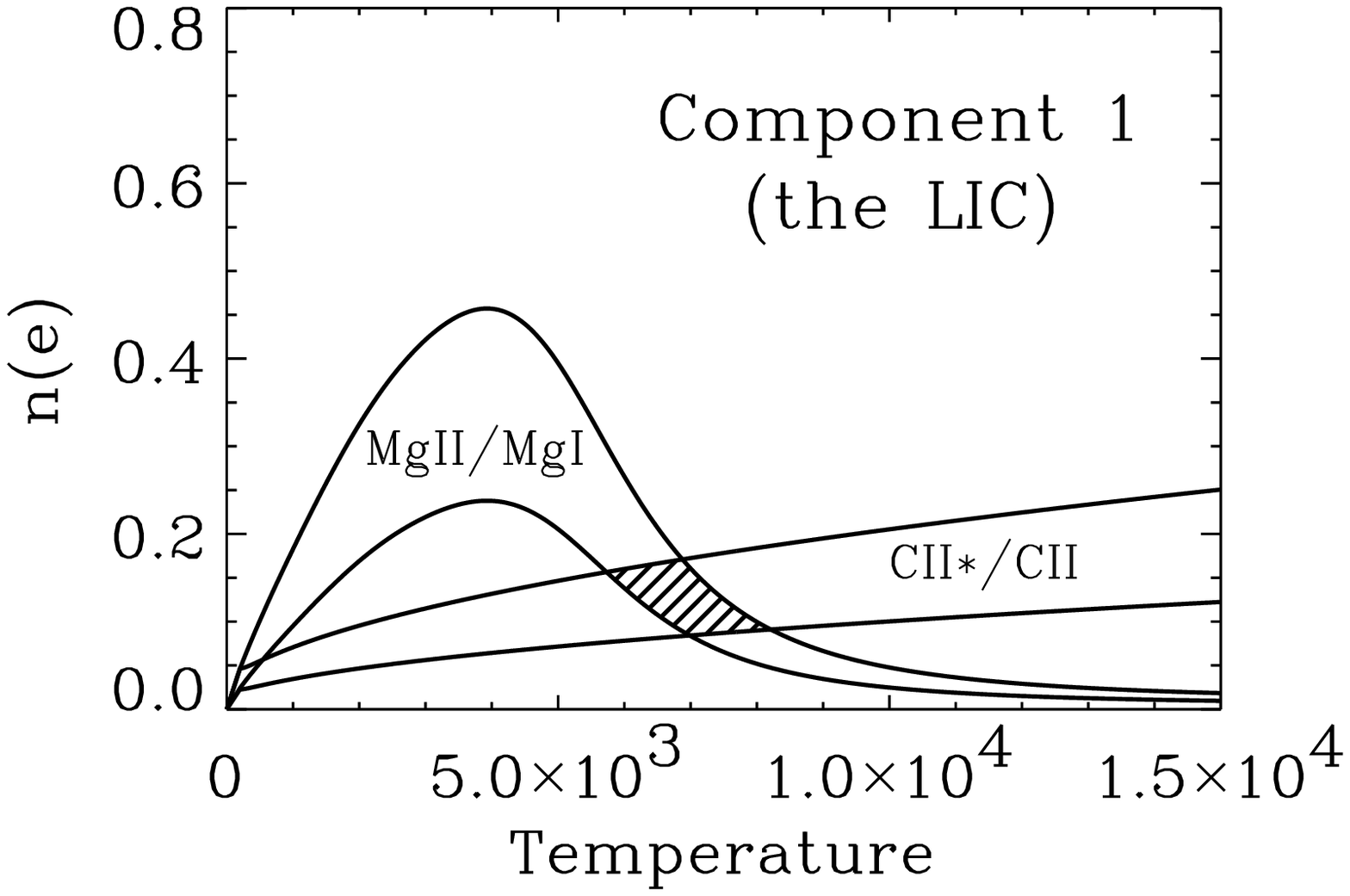,width=5.8cm}\hspace{0cm}
      \psfig{figure=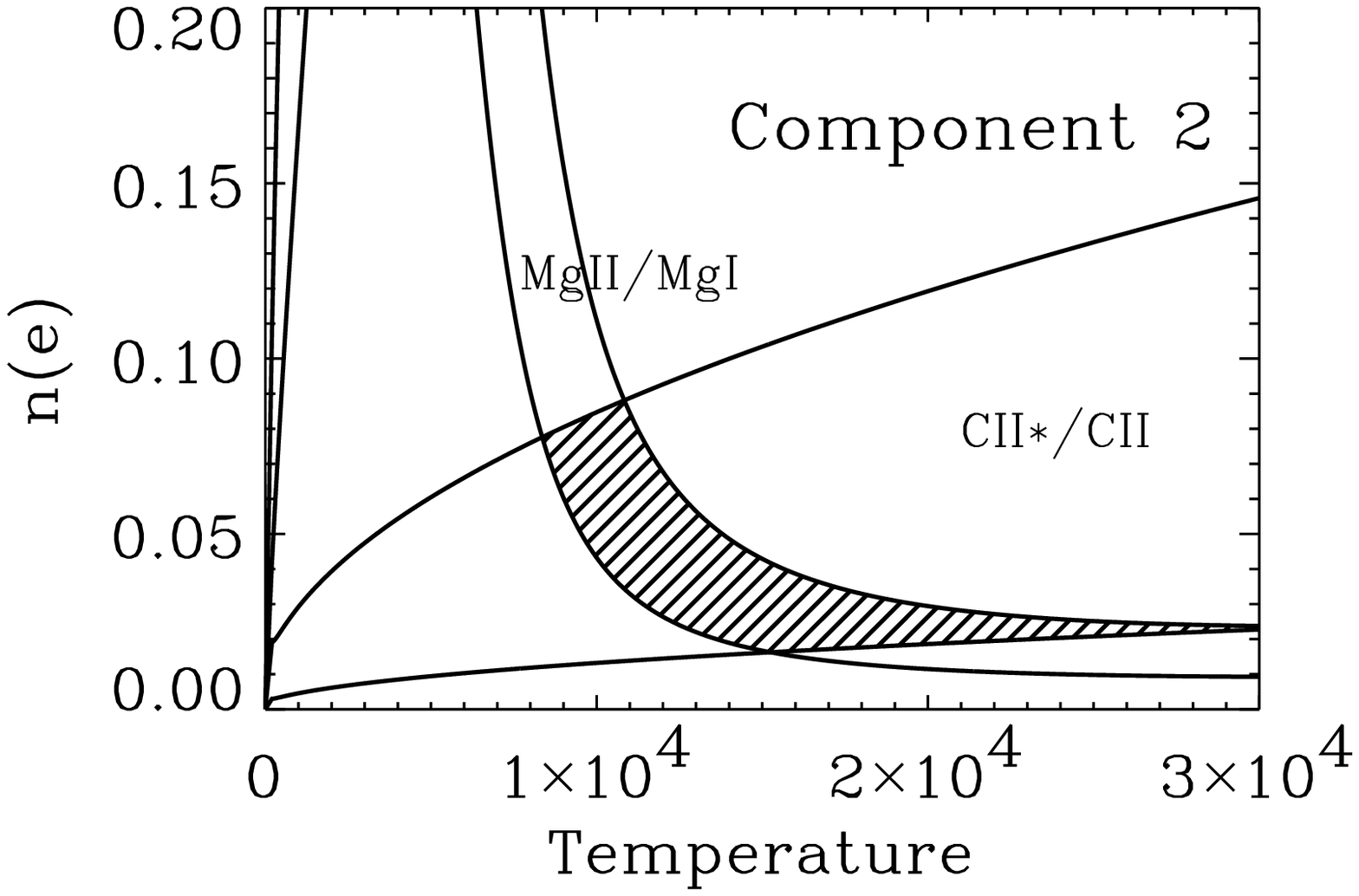,width=5.8cm}\hspace{0cm}
      \psfig{figure=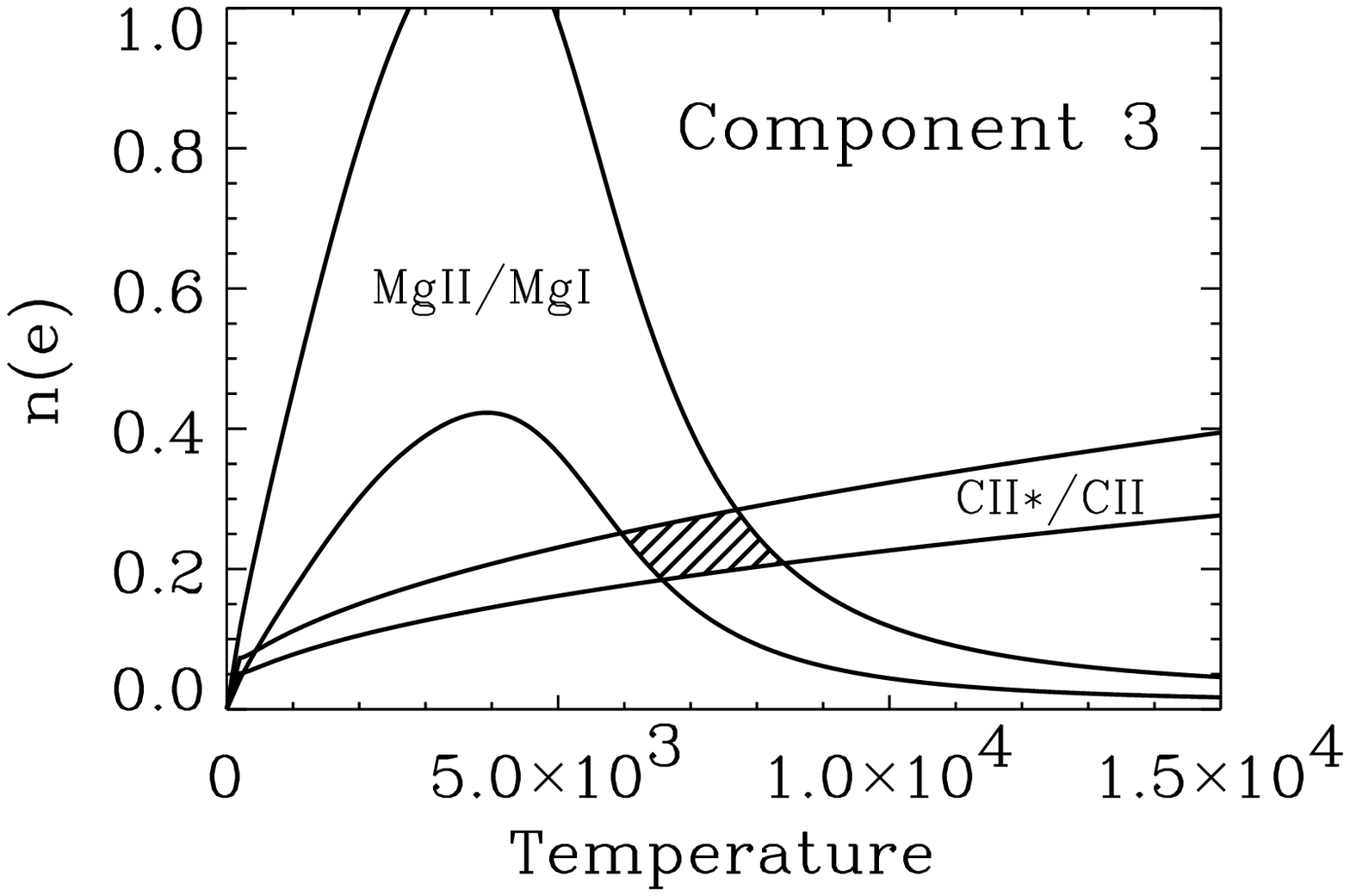,width=5.8cm}}
      \vspace{0cm}
\caption{Curves of the electron density n(e) versus temperature T
            governed by equations ~\ref{n(e)} and \ref{mg_equilib}, for
            the permitted values of the ratios n(\cii*)/n(\cii) and 
            N(\mgii)/N(\mgi). For each of the three components the 
            shaded area 
            where the two sets of curves 
            intersect determines  the possible values for n(e) and T.
    \label{fig:neteps}}
\end{figure*}

We follow the method of Jenkins, Gry \& Dupin (2000) to derive 
both the temperature and the electron density in a component by combining
the information from two ratios which depend 
in different manners on these two physical parameters.
 
The relative populations of the fine-structure levels of \cii\ are
governed by the balance between collisions and radiative de-excitation. In the 
diffuse warm medium, the collisions are dominated by electrons. The 
condition for equilibrium
\begin{equation}\label{equilib}
n(e)\gamma_{1,2}n(\cii)=[n(e)\gamma_{2,1}+A_{2,1}]n(\cii*)
\end{equation}
leads to an equation for the electron density
\begin{equation}\label{n(e)}
n(e)={g_2A_{2,1}T^{0.5}\left[{n(\cii*)\over n(\cii)}\right]\over
8.63~10^{-6}\Omega_{1,2}\left\{ \left({g_2\over g_1}\right) \exp\left(
{-E_{1,2}\over kT}\right) - \left[{n(\cii*)\over n(\cii)}\right]\right\}
}
\end{equation}
where the radiative decay probability for the upper level is
$A_{2,1}=2.29~10^{-6}{\rm s}^{-1}$ (Nussbaumer \& Storey 1981), and the
collision strength $\Omega_{1,2}=2.81$ (Hayes \& Nussbaumer 1984).
Equation~\ref{n(e)} thus gives a curve of $n$(e) versus $T$ for any value
of the ratio n(\cii*)/n(\cii).

To see the conditions from a different perspective, we make use of the equation
for the equilibrium between the two lowest
ionization levels of magnesium which is given by 
\begin{eqnarray}\label{mg_equilib}
\big[\Gamma(\mgi)+C(\mgii)n(H^+)\big]n(\mgi)=&& \nonumber\\
\alpha(\mgi)n(e)n(\mgii)~.&&
\end{eqnarray}
For the charge exchange rate $C(\mgii)$ that applies to the reaction
\mgi~+~H$^+\rightarrow$ \mgii~+~H, we
used the analytical approximation
$C(\mgii)=1.74~10^{-9}\exp(-2.21~10^4/T)$ derived by Allan, et al. (1988). 
We calculate $\Gamma(\mgi)=6.1~10^{-11}{\rm s}^{-1}$ at the position of the Sun
(for details, see Jenkins, Gry \& Dupin (2000)). For
$\alpha(\mgi)$ we used the radiative and dielectronic recombination
rates given by Shull \& van Steenberg (1982), supplemented by the
additional contributions from low-lying resonance states computed by
Nussbaumer \& Storey (1986).  As with Equation~\ref{n(e)}, an application of
Equation~\ref{mg_equilib}
also provides a curve  giving $n$(e)
versus $T$, but this time for any value of the ratio $N$(\mgii)/$N$(\mgi).
 
In Figure~\ref{fig:neteps}, we plot  for each component
the  curve $n$(e)
versus $T$ of Equation~\ref{n(e)} for the maximum  and the minimum values of 
the ratios n(\cii*)/n(\cii) 
derived from the observations, together with the two curves $n$(e) versus $T$
given by Equation~\ref{mg_equilib} for the two extremes for the ratio
$N$(\mgii)/$N$(\mgi).
The shaded area bounded by the two sets of curves defines the possible 
values of $n$(e) and $T$ that are consistent with both methods for each 
component.

We derive $5700<T<8200\,$K with $0.08<n_e<0.17\,$cm$^{-3}$ for the LIC 
(Component~1),
$8200<T<30\,000\,$K with $0.016<n_e<0.088\,$cm$^{-3}$ for Component~2, and
$6000<T<8400\,$K with $0.18<n_e<0.28\,$cm$^{-3}$ for Component~3.
\begin{table}
\caption[]
{Derived properties of the three main components.
}
\begin{tabular}{lccc}
\hline
\noalign{\smallskip}
Comp.&  1 (LIC) & 2 & 3  \\
\noalign{\smallskip}
\hline
\noalign{\smallskip}
T (K) &5700--8200&8200--30\,000&6000--8400\\
\noalign{\smallskip}
n$_e$ (\cmt)  & 0.08--0.17&0.016--0.088&0.18--0.28\\
\noalign{\smallskip}
n$_e$/n$_{tot}$&$<$0.55 &$<$0.43&0.955--0.985\\
\noalign{\smallskip}
n$_{tot}$ (\cmt)  & $>$0.14 & $>$0.034&0.18--0.29\\
\noalign{\smallskip}
p/k (\cmt\,K) & $>$1300 & $>$440 & 2300-5000\\
\noalign{\smallskip}
length (pc)  & $<$1.3 & $<$3.3&0.17--0.45\\
\noalign{\smallskip}
\hline
\end{tabular}
\label{tab:physcond}
\end{table}

Note that  our results confirm the temperature of $7000\,$K commonly
found for the LIC in most studies, and our electron density of 
$n_e=0.12\pm0.05\,$cm$^{-3}$ confirms the ranges found
independently in the LIC in other sight-lines by  Wood \& Linsky (1997) 
and Holberg et al. (1999).
From our results, the LIC and Component~3 have similar temperatures
while Component~2 could be  warmer.
\section{Ionization fractions and  ionization processes}\label{sec:ionfrac}
 From the comparison of the total and the neutral hydrogen column densities 
derived  from N(\sii) and N(\oi) respectively (see Section~\ref{sec:hydrogen}), 
we  estimate the 
ionization fractions n$_e$/n$_{tot}$, listed in Table~\ref{tab:physcond}.

Component~3 is ionized 
by more than 95 \%, making its  ionization comparable to that of the 
two main 
components (C and D) in the line of sight to \bcma\ (Jenkins, 
Gry \& Dupin 2000). As for Components C and D, the main ionizing source for 
Component 3 is 
probably the star \ecma\, and its high hydrogen ionization
fraction requires that it is located closer to the star, further away 
from the Sun.

Component~1
(the LIC) can be at most 55\% ionized, its maximum  ionization 
fraction  corresponding to the case of no oxygen depletion in the local cloud.
However, formally the column density ranges allow the LIC to be neutral.

Since for Component~2 the neutral gas and total gas column density ranges 
are nearly coincident,  Component~2 is most probably neutral although 
formally its ionization fraction can be as high as 43\% in the case 
of no oxygen depletion.

\begin{figure}
      \vspace{0cm}
\psfig{file=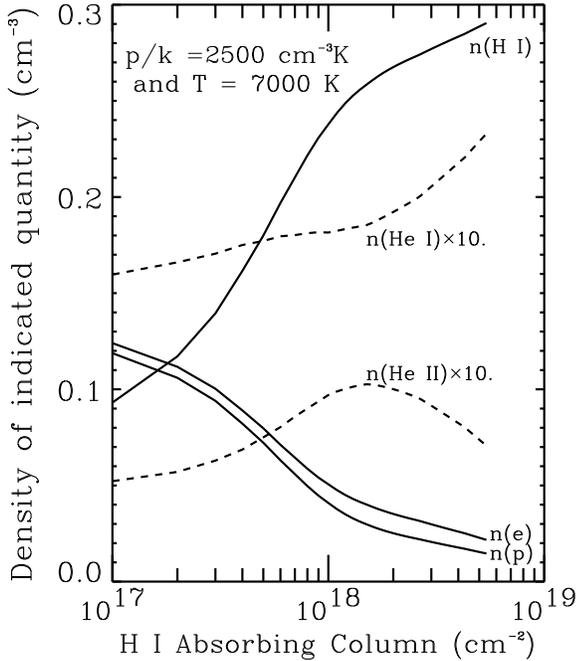,width=8.5cm}
     \vspace{0cm}
\caption{Predicted densities  versus \hi\ attenuation column density in a 
local cloud exposed to 
the local photoionization field (due to  hot stars and 
 hot-gas conductive interface).
\label{fig:model}}
\end{figure}

In order to compare these numbers to the expected ionization fractions,
we have calculated the photoinionization equilibria in the local 
ISM for varying depths of shielding by neutral H and He following the equations 
described by Sofia and Jenkins (1998). We consider the effect of Vallerga's
(1998) composite stellar radiation field supplemented by Slavin's (1989)
calculated flux from the cloud conductive interface. The results are shown 
in Figure~\ref{fig:model}.
 The calculation is the
same as that used to produce Figure~2 of Jenkins et al. (2000) except
for a higher pressure p/k,  in better agreement with the electron density 
derived 
in Section~\ref{sec:density} for the LIC as well as with the neutral density 
derived
from EUV stellar spectra and inside the heliosphere (Vallerga 1996, 
Qu\'emarais et al. 1994).
We derive the expected ionization fraction in the two local 
clouds from 
Figure~\ref{fig:model}
and from the estimate of the mean attenuation of the ionizing flux received in 
the clouds. 
Since most of the 
EUV radiation comes from the direction of \ecma\ where
nearly all the neutral gas  is included in one of these 
two components, the attenuation is derived from their \hi\ column densities
given in Section~\ref{sec:neutral}.

If we assume that Component~2 is located outside of the LIC, further out 
toward \ecma, it
should be  shielded only by its own \hi\ material,
with a mean \hi\ attenuation of $1/2$ N(\hi), in all cases lower than 
2.3~10$^{17}$~\cmd. With this upper limit,
Figure~\ref{fig:model} predicts for Component 2 a minimum ionization 
fraction of 0.44, which is inconsistent with the slightly lower maximum 
ionization fraction permitted by
the observations. Note   that the temperature of Component~2 
derived in Section~\ref{sec:density}
is higher than that used in {Figure~\ref{fig:model},  
optimized to match the characteristics of the LIC. 
As the ionization fraction increases with temperature because the 
recombination rate is a decreasing function of T, 
the predicted ionization fraction for Component~2 is even higher
than that which appears in
Figure~\ref{fig:model}, amplifying further the discrepancy with the value
derived from the observations.
This inconsistency would be eliminated if
Component~2 were shielded by the LIC, which would imply that it  
is located within the Local Cloud, a configuration already proposed 
by Gry (1996) for some of the small components detected close to the Sun in 
several lines of sight. 

For the LIC, the mean attenuation 
due to the cloud itself ranges from 1.5~10$^{17}$ to 2.2~10$^{17}$~cm$^{-2}$
depending on the assumed oxygen 
depletion value. From Figure~\ref{fig:model}, this implies an ionization
fraction between 0.45 and 0.5. If we were to add the attenuation effect
due to Component~2 (which is maximum if Component~2 is external and shielding 
the LIC as a whole), the  range of possible absorbing column is extended up 
to  4 or 6 10$^{17}$~cm$^{-2}$  and we would get an ionization fraction
 down to  0.34 or 0.24.
These values are all compatible with the range derived  for the LIC from the 
measured column densities.

Figure~\ref{fig:model} also predicts that the  ratio n(\hei)/n(\hi) 
is almost constant over the 
range of \hi\ column
densities measured toward the white dwarf stars observed by EUVE 
(around 1\,10$^{18}$ \cmd)  and is close to
the measured ratio of 0.07 (Dupuis et al. 1995). We conclude 
from the adequate prediction of the 
hydrogen ionization fraction as well as of the n(\hei)/n(\hi) ratio that the  
photoionization by  
the EUV radiation field due to the combination of the hot stars and 
the cloud conductive interface  
is a likely representation of the ionization processes in the LIC.
\section{Other physical  properties of the clouds}\label{sec:properties}
From the comparison of the ionization 
fractions  and the electron density, 
we infer  the total hydrogen density n$_{tot}$ (or its lower limit) for 
each of the three 
components. These are listed in Table~\ref{tab:physcond}.
In fact the numbers are compatible
with  all three components having a similar total density  close to
0.2--0.3 \cmt\ but different 
electron densities depending on their
ionization state. 

The thermal pressures in the components are also listed in
Table~\ref{tab:physcond}.
To estimate the thermal pressure in the clouds  from p/k~=~nT we sum up over
all particles : hydrogen (n$_{tot}$=n$_{HI}$+n$_e$), helium (0.1~n$_{tot}$)
and electrons (n$_e$). 
For the LIC, if we adopt  the  n$_{HI}$ range of Vallerga (1996), 
i.e.\ 0.15 to 0.34 \cmt, with our ranges for n$_e$ and T,
we derive  p/k~=~1900 to 6000 \cmt\ K, in
agreement with the thermal pressure found in Component~3, as is expected if
they are in equilibrium with a
surrounding medium that is common to both of them.

The thicknesses of the components
along the line of sight are estimated from the ratio of the total 
column density
N(H$_{tot}$) derived in Section~\ref{sec:totcol} to the total density
n$_{tot}$.  The lengths of
the components (listed in Table~\ref{tab:physcond}) are all very small 
compared to the length of the 
line of sight : they occupy a total of less than 5 pc. This 
implies that at least 96 \% of the sight-line is empty or filled with
more highly ionized gas.

\section{High ionization species and cloud interfaces}\label{sec:interfaces}
Substantial amounts of \siliii\ are detected in all components.
We find that $N(\siliii) = 0.5N(\silii)$ in Component~1 and more than 
$0.7N(\silii)$ in Component~3.  While  $N(\siliii)$ is only about 10\% the
value of  
$N(\silii)$ in Component~2, it is 7.5 times $N(\silii)$ in Component~4.

In principle \siliii\ cannot come from the same
region as the other species  because charge exchange with even small
amounts of neutral hydrogen tends to shift the Si to lower stages of 
ionization
(Jenkins, Gry \& Dupin 2000).  Nevertheless, it seems clear that \siliii\ 
arises at
velocities that coincide with all components, including even the less ionized
Component~2.  This makes it is very
likely that \siliii\ is located in regions associated
with the clouds, perhaps in their outermost layers.

It is interesting
to note that \siliii\ is not detected  
in the spectrum of \acma\ (H\'ebrard et al 1999) up to a limit of 
2~10$^{11}$\cmd.
This limit corresponds to our detection for Component~2, but it is more than
10 times lower than the column density we derive for the LIC. 
This discrepancy could be interpreted as \siliii\
coming from a completely unrelated cloud, which  coincidently has about
the same velocity as the LIC, as proposed by H\'ebrard et al (1999).  

A difficulty with the above proposal is that this
component probably would contaminate other lines,
and thus it should influence the column density determinations for 
other elements. This hypothetical extra component would be likely to have
elemental abundance ratios different from the LIC  --due in particular to 
different ionization fractions-- and thus create large perturbations in the
column density ratios. 
Yet, if we compare the  results
for the four species for which  we have reliable results for both
\acma\ and \ecma\ sight-lines (i.e. derived from unsaturated lines)~: 
\niti, \silii, \feii\ and \mgii, the 
column density ratios between \ecma\ and \acma\ are very similar 
for all elements~: $N($\ecma$)/N($\acma$)=1.5~\pm0.2\,$\cmd.
As an example of the kind of differences one can expect
between two different components, the column density ratios between 
Component~1 and Component~3 for the three
{\it ionized} species \silii, \feii\ and \mgii\  present a dispersion of 60\%.
Even worse,  the \niti\ column density ratio between 
Component~1 and Component~3 is more than a factor of 10
higher than the  ratios for the other elements. 

Thus, in view of the small dispersion in the column density ratios between 
the LIC
in the \ecma\ sight-line and the LIC in the \acma\ sight-line
in the various elements, we strongly support the idea that there is only
one absorbing component at the LIC velocity in the \ecma\ sight-line, and
thus that the \siliii\ absorption is related to the LIC component.

The fact that N(\siliii) is at least a factor of 10 lower  in the Sirius 
sight-line than in the \ecma\ sight-line suggests
that the region where the \siliii\ originates is extended and that more than
90\% of it lies beyond Sirius. The possible very small
velocity shift
between the \siliii\ absorption and the absorption from less ionized species
could be explained by the presence of a slight positive velocity gradient
toward the outer layer of the cloud.

The detection of  \siliii\ in the LIC has an interesting
consequence for the relative location of the two local clouds. 
Since by definition  the LIC is the cloud in which the Sun is
embedded,  the presence of an extended LIC \siliii\ layer past Sirius
implies that the LIC and its extended layer occupy the full line of sight
toward Sirius. It follows that Component~2 
(the ``Blue Component'' in the \acma\ sight-line)
should be embedded in the LIC or at least in its extended \siliii\ 
layer, corroborating the suggestion made in Section~\ref{sec:ionfrac}.

There is a significant difference between the gas responsible for the 
\siliii\ absorption and the gas responsible for the \civ\ absorption~:
while the widths of the \siliii\ profiles are compatible with \siliii\
being at the same temperature as the less ionized elements, the profiles for
\civ\ are clearly  broader, implying  temperatures of the order of
100\,000 to 200\,000\,K.
This favours the existence of collisional ionization due to a high
temperature. Indeed, our \civ\ column densities and
our \siliv\ upper limits are compatible with the outcome from Slavin's (1989)
model for the conduction layer between the Local Cloud and the hot gas
that is supposed to fill the Local Bubble.  He calculated that
$N$(\civ)~=~2.7~$10^{12}$\cmd\ and $N$(\siliv)~=~1~$10^{11}$\cmd\ through
an interface of this sort, assuming minimal inhibiting effects from
magnetic fields.  His prediction for $N$(\siliii) of $5~10^{10}$\cmd\ is
far below our observed column density.
\section{Summary}\label{sec:summary}
We have analysed the interstellar absorption lines in the high spectral 
resolution ($R\sim100\,000$) UV spectrum
of \ecma\ (130~pc).
We derive column densities for 11 different elements in the three main clouds
and for a few elements with the strongest lines in two additional
very weak components. 

Two of the main components (Components~1 and 2) are identified with the two
components detected in the much shorter line of sight
toward \acma\ (Sirius) which is not far from \ecma\ in the sky. 
One of them (Component~1) is the Local Interstellar Cloud (LIC)
in which the Sun is embedded. For the four elements for which reliable 
measurements exist for the two lines of sight (i.e. performed with unsaturated
lines) we find a constant ratio of 1.5 between the column densities of the LIC
toward \ecma\ and  the column densities of the LIC toward \acma.

We derive the neutral hydrogen column density from our measurement of \oi\ 
which is a good tracer of \hi.
Depending on the oxygen abundance we  adopt, we find a neutral gas
column density for the whole line of sight between 6.0$\pm$1.2~10$^{17}$\cmd\
if we consider that the local gas is not depleted (adopting the B stars 
abundance)
and  9.0$\pm$2.0~10$^{17}$\cmd\ if we adopt the mean ISM oxygen abundance of
Meyer et al. (1998).
With the same two alternatives, we derive for the LIC a neutral gas
column density between  3.0$^{+1.1}_{-0.4}$~10$^{17}$\cmd\ and
4.4$^{+1.6}_{-0.6}$~10$^{17}$\cmd.

We estimate the temperatures and electron densities in the three main
components by combining the information of the two ratios N(\cii*)/N(\cii)
and N(\mgii)/N(\mgi). In particular for the LIC we find  
\dens\ = $0.12\pm0.05$ \cmt\ and T = 7000$\pm1200$~K, both in  agreement
with previous determinations having similar error bars.

We compare the  neutral gas column densities with the total (neutral 
and ionized) gas column densities derived from the \sii\ measurements,
to conclude that Component~3 is ionized by more than 95~\%, that Component~2 
is probably  neutral but could be as much as 43\% ionized and that the LIC 
can be at most 55~\% ionized. 
We conclude from
these numbers that Component~3 must be located further away on the line of 
sight and is
thus almost fully ionized by \ecma\ and that the ionization fraction in the 
LIC is compatible with the gas
being  ionized by the local EUV radiation 
fields from the hot stars and the cloud interface with hot gas. 
In contrast,
it is hard to explain the low state of ionization of Component~2 unless
it is included within the LIC which shields it from the ionizing radiation.

We detect high ionization species. \siliii\ is detected in all clouds
but more significantly in the LIC and Component 3, and is probably located in
extended layers in the outer regions of the clouds. \civ\ is also
detected in the LIC and Component 3, but with small velocity offsets from the
lower ionization species. The derived amount of highly ionized gas and 
the derived high temperature are consistent 
with the predictions of  Slavin (1989) for a conductive interface between
the Local Cloud and the surrounding hot  gas from the Local Bubble.

%________________________________________ Do not leave a blank line here!
%
\begin{acknowledgements}
The GHRS data reduction and a preliminary spectral analysis 
 have been performed in collaboration with Olivier Dupin as part of his 
PhD thesis, presented in April 1998.\\
CG is very grateful to Martin Lemoine for his absorption line fitting
software 'Owens' and his helpful advices.\\
EBJ was supported by NASA grant NAG5-616 to Princeton University.
\end{acknowledgements}


\begin{thebibliography}{}
\bibitem{} Allan R.J., Clegg R.E.S., Dickinson A.S., Flower D.R., 1988,
MNRAS, 235, 1245
\bibitem{} Anders E. and Grevesse N.,
1989, Geochim. Cosmochim. Acta, 53, 197
\bibitem{}  Cassinelli J.P., Cohen D.P., MacFarlane J.J., Drew J.E., 
Lynas-Gray A.E., Hoare M.G., Vallerga J.V., Welsh B.Y., Vedder P.W., 
Hubeny I., Lanz T., 1995, ApJ. 438, 932
\bibitem{}  Dupin O., Gry C., 1998,
A\&A 335, 661
\bibitem{} Dupuis J., Vennes S., Bowyer S., Pradhan A., Thejll P., 1995,
 ApJ 455, 574 
\bibitem{}  Field, G. B. and Steigman, G. 1971 : ApJ 166, 59
%\bibitem{}  Fitzpatrick E.L.,  1996, ApJ 473, L55
\bibitem{}  Fitzpatrick E.L. and Spitzer L., 1997 ApJ 475, 623
\bibitem{} Frisch P.C. and Slavin J.D. 1996, Space Sci. Rev. 78, 223
\bibitem{}  Gry C., Lemonon L., Vidal-Madjar A., Lemoine M., Ferlet R.,
 1995, A\&A 302, 497
\bibitem{} Gry C., 1996, Space Sci. Rev., 78, 239
\bibitem{} Hayes M.A., Nussbaumer H., 1984, A\&A 134, 193
\bibitem{} H\'ebrard G., Mallouris  C.,  Ferlet R.,  Koester D., Lemoine M., 
Vidal-Madjar  A. and  York D.G., 1999 A\&A 350, 643
\bibitem{} Hoffleit D., and Jaschek C. 1982, The Bright Star 
Catalogue,  4th ed., (New Haven: Yale U. Obs.)
\bibitem{} Holberg J.B., Bruhweiler F.C., Barstow M.A., Dobbie P.D., 1999,
ApJ 517, 841
\bibitem{} Hurwitz M., Bowyer S., Bristol R., Dixon W.V., Dupuis J., Edelstein J.,
Jelinsky P., Sasseen T.P., Siegmund O., 1998, ApJ 500, L1
\bibitem{} Jenkins E.B., Reale M.A., Zucchino P.M., Sofia U.J., 1996,
           Astr. and Space Sci. 239, 315
\bibitem{} Jenkins E.B., Gry C. and Dupin O., 2000, A\&A 354, 253
\bibitem{} Jenkins E.B., Oegerle W.R., Gry C., Vallerga J., Sembach K.R.,
Shelton R.L., Ferlet R., Vidal-Madjar A., York D.G., Linsky J.L., Roth K.C.,
Dupree A.K., Edelstein J. 2000: ApJ 538, L81
\bibitem{}  Lallement R. and Bertin P.,
1992, A\&A 266, 479
\bibitem{}    Lallement R., Bertin P.,  Ferlet R., Vidal-Madjar A.,
Bertaux J.L.,
1994, A\&A 286, 898.
\bibitem{} Lallement R. and Ferlet R., 1997, A\&A 324, 1105
\bibitem{} Linsky J.L., Diplas A., Wood B.E., Brown A., Ayres T.R.,
 Savage B.D., 1995: ApJ 451, 335
\bibitem{} Lyu C.H. and Bruhweiler F.C., 1996, ApJ 459, 216
\bibitem{} Meyer D.M., Cardelli J.A., Sofia U.J., 1997, ApJ 490, L103
\bibitem{} Meyer D.M., Jura M., Cardelli J.A., 1998, ApJ 493, 222
\bibitem{}  Morton D.C., 1991, ApJS 77, 119
\bibitem{} Nussbaumer H., Storey P.J., 1981, A\&A 96, 91
\bibitem{} Nussbaumer H., Storey P.J., 1986, A\&AS 64, 545
\bibitem{} Perryman M. A. C., Lindegren L., Kovalevsky J., Hog E.,
 Bastian U., Bernacca P. L., Cr\`eze M., Donati F., Grenon M., Grewing 
M., van Leeuwen F., van der Marel H., Mignard F., Murray C. A., Le Poole 
R. S., Schrijver H., Turon C., Arenou F., Froeschl\'e M., and Petersen C. 
S. 1997, A\&A  323, L49
\bibitem{} Qu\'emarais E., Bertaux J.L., Sandel B., Lallement R., 1994,
A\&A 290, 941
\bibitem{} Reynolds R.J., 1986, AJ 92, 653
\bibitem{} Savage B.D. \& Sembach K.R., 1996, ARAA 34, 279
\bibitem{} Sembach K.R., Savage B.D., Jenkins E.B., 1994, ApJ 421, 585
\bibitem{} Shull J.M., van Steenberg M., 1982, ApJS 48, 95
\bibitem{}  Slavin J., 1989, ApJ 346, 718.
\bibitem{} Slavin J.D. and Frisch P.C., 1998, in The Local Bubble and Beyond, 
ed. Breitschwerdt, Freyberg and Trumper (Berlin: Springer), p305
\bibitem{} Sofia U.J. and Jenkins E.B., 1998, ApJ 499, 951 
\bibitem{}  Soderblom 
D.R., Gonnella A. , Hulbert S.J.,
 Leitherer C., Schultz A., Sherbert L.E., 1995, GHRS Instrument Handbook 
 version 6, Space 
 Telescope Science Institute.
\bibitem{} Vallerga J.V., 1996,  Space Sci. Rev., 78, 277
\bibitem{} Vallerga J.V., 1998, ApJ 497, 921
\bibitem{} Vallerga J.V. and Welsh B.Y., 1995, ApJ 444, 702
\bibitem{}  Vallerga J.V., Vedder P.W., Welsh B.Y.,
1993, ApJ 414, L65
\bibitem{} Wood B.E. and Linsky J.L., 1997, ApJ 474, L39 
\end{thebibliography}
\end{document}